\documentclass[12pt]{iopart}
\usepackage{graphicx}
\usepackage{caption}
\usepackage{multirow}
\usepackage{varioref}

\usepackage{hyperref}

\usepackage{color}

\graphicspath{ {./Figures/} }

\begin{document}

\title[UV LED Lifetime Testing]{Lifetime Testing UV LEDs for Use in the LISA Charge Management System}

\author{D Hollington, J T Baird, T J Sumner and P J Wass}

\address{High Energy Physics Group, Physics Department, Imperial College London, Prince Consort Road, London, SW7 2BW, UK.}
\ead{d.hollington07@imperial.ac.uk}

\begin{abstract}

As a future charge management light source, UV light-emitting diodes (UV LEDs) offer far superior performance in a range of metrics compared to the mercury lamps used in the past. As part of a qualification program a number of short wavelength UV LEDs have been subjected to a series of lifetime tests for potential use on the Laser Interferometer Space Antenna (LISA) mission. These tests were performed at realistic output levels for both fast and continuous discharging in either a DC or pulsed mode of operation and included a DC fast discharge test spanning 50 days, a temperature dependent pulsed fast discharge test spanning 21 days and a pulsed continuous discharge test spanning 507 days. Two types of UV LED have demonstrated lifetimes equivalent to over 25 years of realistic mission usage and provide a baseline and backup solution for LISA.

\end{abstract}


\section{Introduction}

The free-falling test masses within the proposed space-based gravitational-wave observatory LISA \cite{Audley2017} will inevitably become electrically charged due to incident ionising radiation from space \cite{Jafry1996}. Left unchecked, such charging will limit the performance of the instrument as the charged test masses can interact with the electric fields within the sensor as well as any magnetic fields present \cite{Sumner2000, Weber2012, Armano2017}.

The technology demonstrator mission LISA Pathfinder \cite{McNamara2008} incorporates a non-contact Charge Management System (CMS) to counter this problem \cite{Sumner2009}. It uses $253.7\,\textup{nm}$ ultraviolet light emitted by low-pressure mercury lamps to generate photoelectrons from both the gold-coated surfaces of the test mass as well as the surrounding housing. By controlling these opposing photoelectric currents the test mass can be charged positively or negatively. However, this simple idea is made more complicated by illuminated surfaces being highly reflective, the possibility of mismatched photoelectric yields as well as by the applied AC and DC electric fields present within the sensor \cite{Schulte2009, Ziegler2014}.

Since the development of the LISA Pathfinder CMS, UV LEDs have become commercially available which offer many advantages over traditional mercury lamps. These include mass, volume and electrical power savings, faster response times and the possibility of using light of a shorter wavelength. Extensive testing of such devices has demonstrated their suitability for use within a future CMS with tests including general electrical and UV output power measurements, spectral stability, pulsed performance, temperature dependence as well as thermal vacuum, radiation and vibration survivability \cite{Hollington2015, Olatunde2015, Saraf2016}.

Device lifetimes under realistic operational scenarios for LISA remain unclear. Within the literature there is evidence that devices with peak wavelengths of $255\,\textup{nm}$ supplied by Sensor Electronic Technology Inc. (SETi) \cite{SETi2015} have lifetimes of several years when pulsed at $1\,\textup{or}\,20\,\textup{kHz}$ with a $10\%$ duty cycle and $2\,\textup{mA}$ drive current amplitude \cite{Sun2009}. However, this is not necessarily how such devices will be operated in the LISA charge management system. Furthermore, it may be desirable for a potential discharging light source to produce at least some light with a wavelength less than $237.4\,\textup{nm}$, which is equivalent to the work function of pure gold deposited in vacuum \cite{Huber1966}. Above this wavelength some surface contamination (predominately from water and possibly some hydro-carbons \cite{Schulte2009, Hechenblaikner2012}) is required to lower the gold work function. While such contamination inevitably occurs due to the surfaces being exposed to air during integration it also seems to persist once the system is placed under vacuum and baked \cite{Hechenblaikner2012}. LISA Pathfinder has a residual pressure within the sensors which limits its sensitivity between $0.7$ and $20\,\textup{mHz}$ \cite{Armano2016}. It is possible that for LISA a more aggressive bake-out procedure will be pursued in order to reduce the final pressure within the sensors. Though unlikely that the surfaces would ever be completely cleaned, using a lower wavelength device for discharging would remove any fear that such a scheme could clean the gold surfaces ``too well'' and therefore prevent charge control.

The aim of this work was to carry out lifetime tests under a variety of realistic operational scenarios on the lowest wavelength devices identified in \cite{Hollington2015}.


\section{Operational Scenarios for LISA}

Charging for LISA will predominately be caused by Galactic Cosmic Rays (GCR) which consist primarily of protons, with around 12\% helium nuclei, 1\% heavier nuclei and 2\% electrons \cite{Simpson1983}. This flux of high energy particles incident on the spacecraft acts to charge the test masses positively at rates of a few +10's of elementary charges per second, dependent on the solar cycle and the voltages within the sensor \cite{Armano2017, Araujo2005}. In addition, Solar Energetic Particle (SEP) events are predicted to be a significant but transient source of test-mass charging for LISA. Such events last from hours to several days and simulations predict charging rates can be enhanced by several orders of magnitude compared to the GCR background rate \cite{Wass2005}. Not all SEP events produce additional charging; only particles with energies greater than $\approx100\,\textup{MeV/n}$ are energetic enough to reach the shielded test masses. SEP events with such particle energies typically occur up to a few times a year at solar maximum but are less frequent during solar minimum \cite{Araujo2005}. The final source of test mass charging has a local origin. Due to a difference in work function, when the retractable plungers that position the test masses disengage they can leave a residual charge. This not only occurs during the initial release but also if the experiment is forced to enter a safe mode when the TMs must be secured mechanically. A residual charge of up to $10^{8}\,\textup{elementary charges}$ is possible, equivalent to  a test mass potential of about $0.5\,\textup{V}$.

\subsection{DC Discharging}

For LISA Pathfinder, there are broadly three different methods used to counter test mass charging and these will be referred to as DC discharge modes. Possibly the simplest mode is called DC fast discharge. Within this open loop scheme the test mass charge is allowed to build up for a period of time based on the observed environmental charge rate until a predetermined threshold is reached. For GCR charging, this period is typically 1-3 weeks. During a SEP event, however, discharging may be required on time-scales of hours and after test mass release it could be necessary immediately. For a DC fast discharge the UV light source is turned on at a relatively high output level for several minutes in order to bring the test mass charge back to a desired level as quickly as possible and allow science operations to be resumed. If necessary DC voltages can be applied within the sensor temporarily to shift the test mass potential with respect to the housing such that any photo-current in the undesired direction is suppressed, enhancing the discharge rate.

Alternatively, DC continuous discharge aims to keep the test mass charge permanently close to zero, minimising the coupling to noisy stray fields in the sensor. The sensor is continuously illuminated to obtain a discharging rate which matches the environmental charging rate, but with the opposite polarity. The apparent yield is the net change in test mass charge per photon entering the sensor. It can be positive or negative and depends on the test-mass potential. Assuming that the background charging rate is positive, continuous discharge requires a negative apparent yield when the test-mass potential is close to $0\,\textup{V}$. A low UV power setting is chosen that results in a net negative photo-current that counters the positive background charging rate. The success of such an approach depends strongly on the light distribution, electrostatics and photo-emission properties of the system which can vary in different part of the sensor. 

A different approach to DC fast or continuous discharge involves exploiting the test mass equilibrium potential. For a particular illumination there is a test mass potential where the apparent yield goes to zero, i.e. where the photo-current in both directions balance. If the equilibrium potential is close to $0\,\textup{V}$ simply turning on the UV light source at a setting high enough to dominate over the background charging rate results in a DC continuous discharge. A caveat to this approach is that DC biases may need to be applied to shift the equilibrium potential in the likely event that it is not exactly at $0\,\textup{V}$. However this is not always possible depending on the properties of the overall system. In addition, care is required such that the UV power is not set so high that it introduces excess noise by transferring large amounts of charge despite the net being zero.

\subsection{Pulsed Discharging}

The use of UV LEDs opens up the possibility of a new operational scenario, so-called pulsed discharging \cite{Sun2006}. By synchronising the UV emission with an existing high frequency AC voltage within the sensor (sensing or actuation), one can ensure illumination is only on when the electric field is favourable for discharge in the desired direction. At the same time any photo-current in the opposite (undesired) direction would be suppressed. This concept has several advantages over DC discharging with the caveat of increased complexity and that it is dependent on most of the absorbed light being in a region that the electric fields can influence. It is particularly appealing if the housing and test mass surfaces have significantly asymmetric quantum yields and also provides redundancy as shifting the phase of the pulses by $180^{\circ}$ would reverse the direction of net photo-current. Another advantage is that there is scope for a very high discharging rate dynamic range which could be varied by adjusting the phase, duration or amplitude of the UV pulses \cite{Ziegler2014}.

At UV photon energies around $4.9\,\textup{eV}$, air-exposed gold surfaces typically produce photoelectrons with energies less than $1\,\textup{eV}$. To have a significant influence on photoelectron flow, a suitable AC voltage for pulsed discharging therefore needs an amplitude of over $~1\,\textup{V}$. In the likely event that LISA will employ a similar sensing and actuation scheme to that of LISA Pathfinder, the obvious AC voltage within the sensor to synchronise to would be the $100\,\textup{kHz}$ injection voltage. This $\pm4.88\,\textup{V}$ bias is applied to six of the electrodes within the sensor and used for capacitive sensing of the test mass \cite{Weber2003}. It also introduces a sinusoidal potential difference between the test mass and grounded housing of $\pm0.6\,\textup{V}$. Since the potential difference between TM and electrode has the opposite sign to that between TM and ground, it would be best to either illuminate the injection electrode regions exclusively ($\pm4.88\,\textup{V}$) or alternatively the regions away from the direct influence of all electrodes ($\pm0.6\,\textup{V}$).

Analogous to the DC operational modes introduced previously, there are then two forms of pulsed discharging. The net discharging rate can be relatively high in order to perform a so called pulsed fast discharge or chosen to match the charging rate, so called pulsed continuous discharging.


\section{Lifetime Testing}

Testing was designed to demonstrate the UV LEDs capability of being operated in a LISA-like DC fast discharge mode, a pulsed fast discharge mode and a pulsed continuous mode. The general philosophy was to push the devices to operational extremes while still staying within data sheet recommendations. In the same way as described in \cite{Hollington2015}, a set of reference measurements were made with every device before and after each lifetime test. Each set of reference measurements were made with the device held in a temperature controlled mount at $20\,^{\circ}\textup{C}$ and consisted of three parts: a current-voltage scan between $0\,\textup{mA}$ and $10\,\textup{mA}$ while the total UV output power was simultaneously recorded, a spectral scan with the device driven at $1\,\textup{mA}$ DC and a waveform measurement while the device was driven at $100\,\textup{kHz}$, $10\%$ duty cycle with a $10\,\textup{mA}$ drive current amplitude.

\subsection{Test Devices}

Two types of device underwent lifetime testing. The first was supplied by Sensor Electronic Technology Inc. (SETi) with part reference UVTOP-240-TO39-HS \cite{SETi2015}. Each device was packaged in a hermetically-sealed TO-39 can with an integrated hemispherical lens and had a nominal peak wavelength of $240\,\textup{nm}$. The nine devices tested here had measured peak wavelengths of $247.2\pm0.2\,\textup{nm}$, just within the limits stated by the manufacturer (maximum peak wavelength of $250\,\textup{nm}$) and spectral FWHMs of $10.1\pm 0.2\,\textup{nm}$. Around 85\% of the light produced by these particular devices was at a wavelength less than $253.7\,\textup{nm}$ (mercury spectral line used in LISA Pathfinder CMS), a wavelength low enough to generate photoelectrons from a gold surface that has been exposed to air. Note also that approximately 1\% of the light should be of a wavelength low enough to liberate photoelectrons from atomically clean gold ($237.4\,\textup{nm}$) \cite{Huber1966}.

\begin{figure}[h]
\begin{minipage}[t]{0.5\textwidth}
\centering
\includegraphics[width=1.0\textwidth]{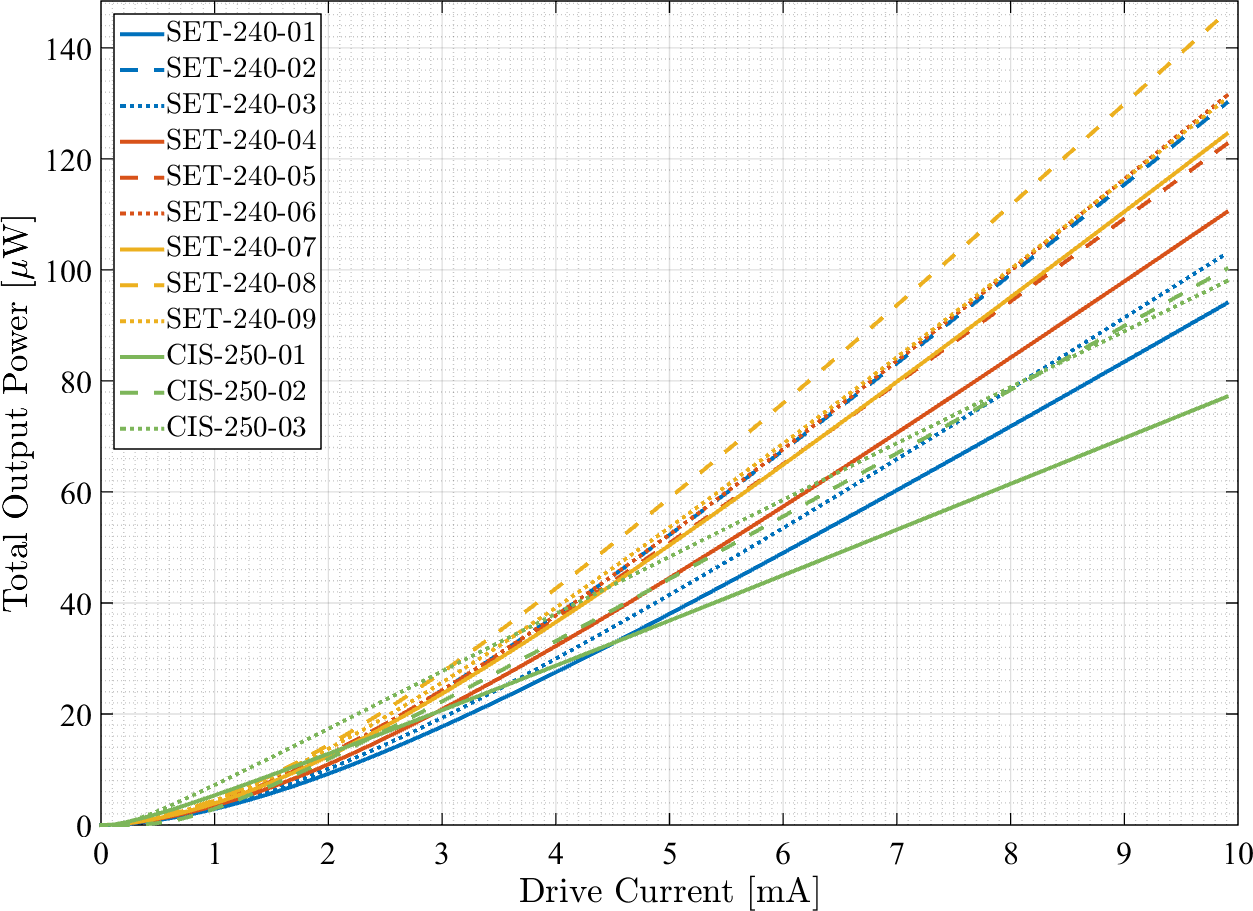}
\includegraphics[width=1.0\textwidth]{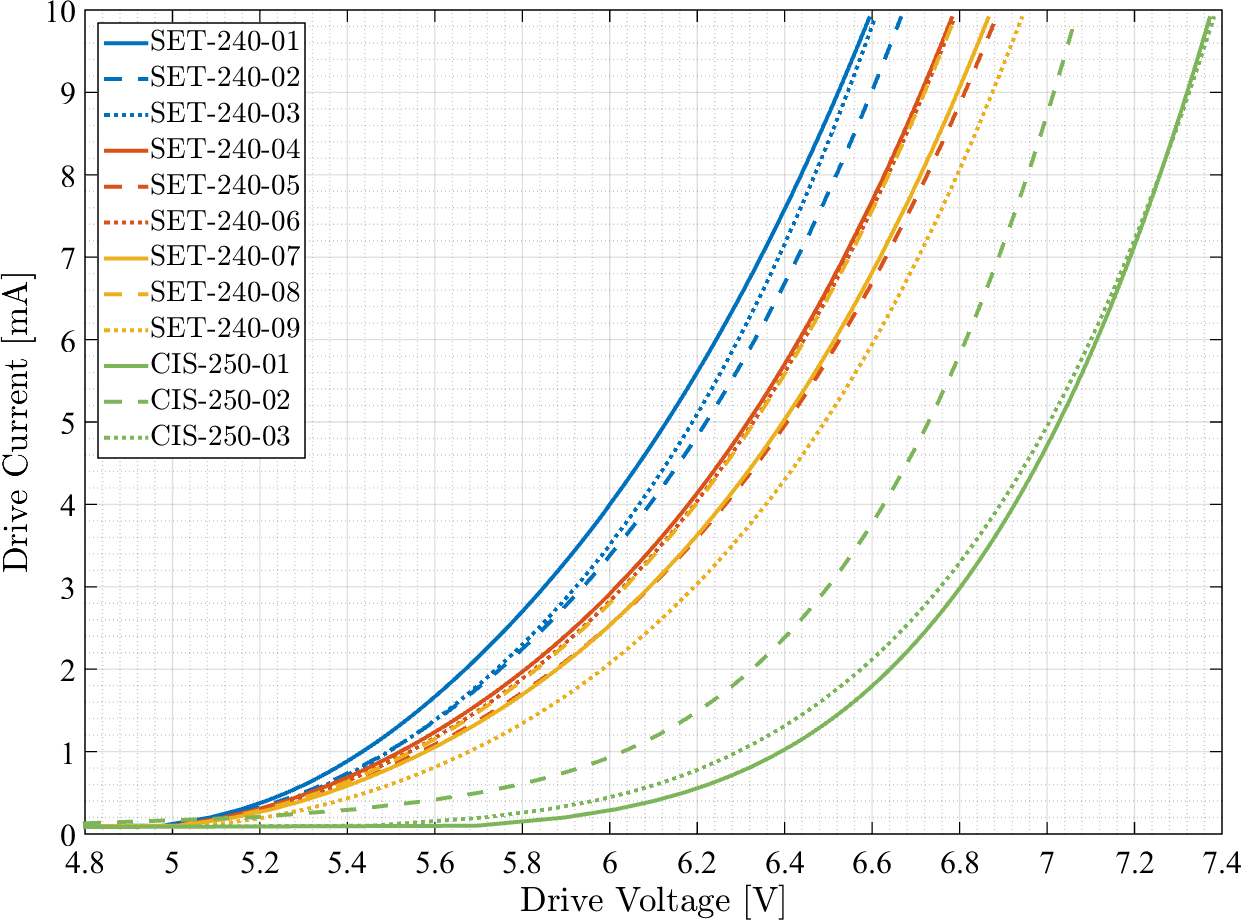}
\end{minipage}
\begin{minipage}[t]{0.5\textwidth}
\includegraphics[width=1.0\textwidth]{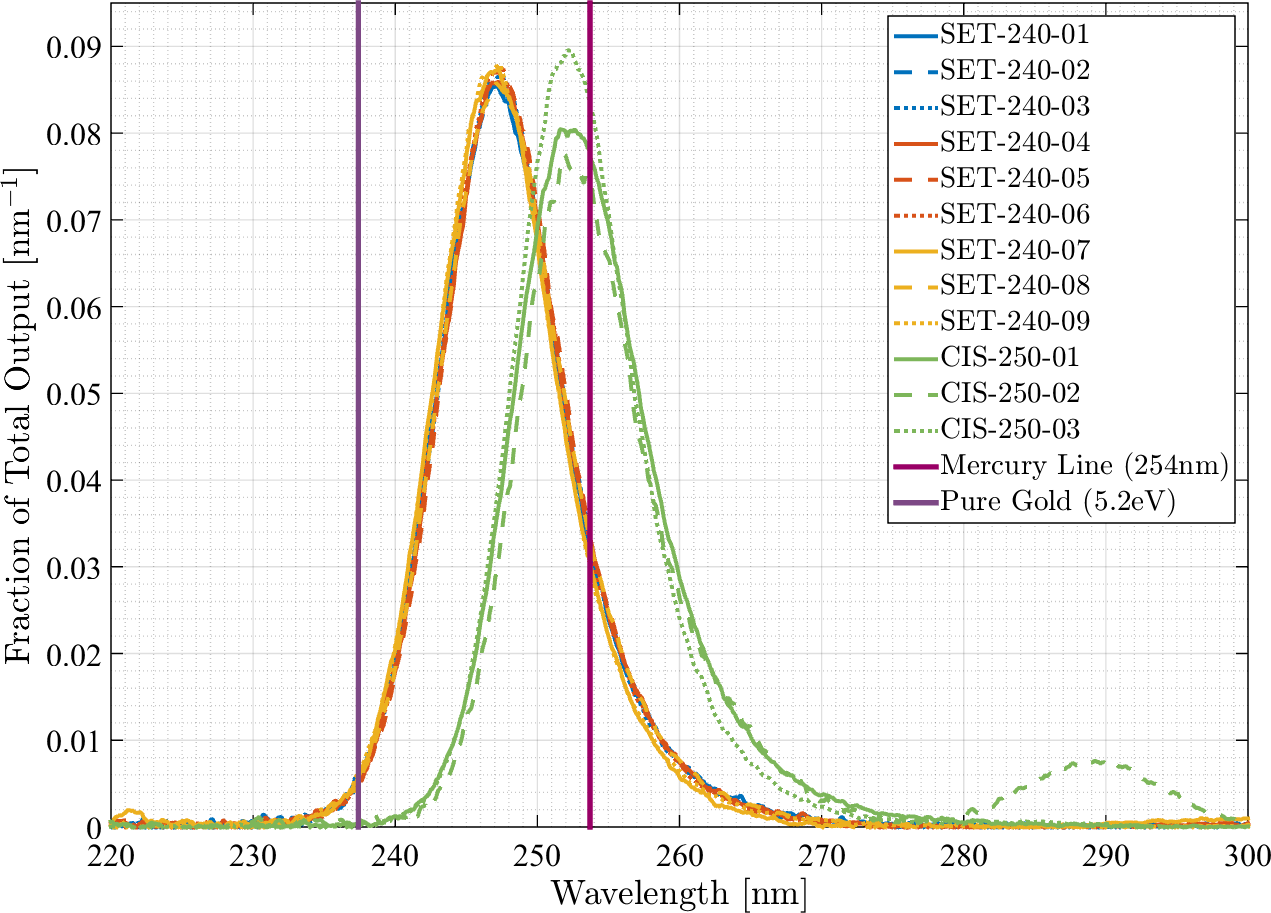}
\caption[Initial device characteristics.]{\label{fig:DeviceChar} Clockwise from bottom left: IV scans, simultaneously measured total UV output power and spectral scans. Note the SET-240 spectra are so similar that they appear almost indistinguishable here. The CIS-250 devices show more variability, with CIS-250-02 even producing an unexpected secondary peak at $289\,\textup{nm}$.}
\end{minipage}
\end{figure}

The second type of device was supplied by Crystal IS (CIS) with a part reference OPTAN250J \cite{CIS2015}. These devices were packaged in hermetically-sealed TO-39 cans with integrated ball lenses and had nominal peak wavelengths of $250\,\textup{nm}$. Three devices of this type were tested and were found to have measured peak wavelengths of $252.4\pm0.2\,\textup{nm}$ and spectral FWHMs of $10.2\pm 0.6\,\textup{nm}$. Just over half the light that these devices produce was at a wavelength less than the $253.7\,\textup{nm}$, but no light was generated at $237.4\,\textup{nm}$ and below. Unlike the SET-240 devices the CIS-250s undergo a $48\,\textup{hour}$ “burn-in” at $100\,\textup{mA}$ prior to delivery. Likely due to this, previous testing has shown CIS-250 devices to have a very consistent performance \cite{Hollington2015}. The pre-testing properties of all twelve devices are collated in Figure \ref{fig:DeviceChar}.

\subsection{DC Fast Discharge}

This test was designed to simulate a simple DC fast discharge scenario for a mission lasting multiple years. It consisted of each device being turned on for 240 seconds (the fast discharge) then turned off for 240 seconds (a greatly reduced rest time between fast discharges). This was repeated for 9000 cycles meaning the total test lasted 50 days.

The set up consisted of a temperature-controlled copper mount held at $20\,^{\circ}\textup{C}$, with space for three devices to be operated simultaneously with independent photodiodes (Burr-Brown OPT301) mounted opposite offering \textit{in situ} output readings. The separation distance between the devices and photodiodes was $25\,\textup{mm}$ with the sensitive photodiode area only measuring a fraction of the emitted light (about 5\%). The copper mount itself was sealed in a light-tight plastic enclosure with data acquisition and system control handled autonomously by a dedicated desktop computer.

The test was carried out twice, once with a group of three SET-240 devices and then repeated with a group of three CIS-250 devices. The actual devices tested were the same six (01, 02 and 03 of each type) that had previously undergone the full performance tests described in \cite{Hollington2015}. Each device was driven independently at a fixed current with amplitudes from $4.6\,\textup{mA}$ to $8.5\,\textup{mA}$ depending on the particular device but chosen to provide a prescribed common UV output at levels required for a pessimistic discharge scenario, derived below.

Assuming a low quantum yield of $10^{-7}$ charges per absorbed photon for a gold surface that had been exposed to air, it would take $\approx10^{12}$ photons with a wavelength below $254\,\textup{nm}$ entering the sensor to discharge a test mass with a charge of $10^{7}\,\textup{e}$ in the chosen $240\,\textup{seconds}$. Based on their measured characteristics (Figure \ref{fig:DeviceChar}) the SET-240 devices emit 87\% of their light below $254\,\textup{nm}$ while the CIS-250 only 55\%. Pessimistically, as little as 10\% of the light emitted by a device might actually reach the sensor via the fibre-optic routing system. Taking these factors into account, drive current levels were chosen to achieve the required number of photons at the sensor and equated to a total UV output power of $50\,\mu\textup{W}$ for the SET-240 devices and $65\,\mu\textup{W}$ for the CIS-250 devices. For context, at their highest setting the mercury lamps employed for LISA Pathfinder produce approximately $7\,\mu\textup{W}$ at the end of their optics barrel.

\begin{figure}[h]
\centering
\includegraphics[width=0.32\textheight]{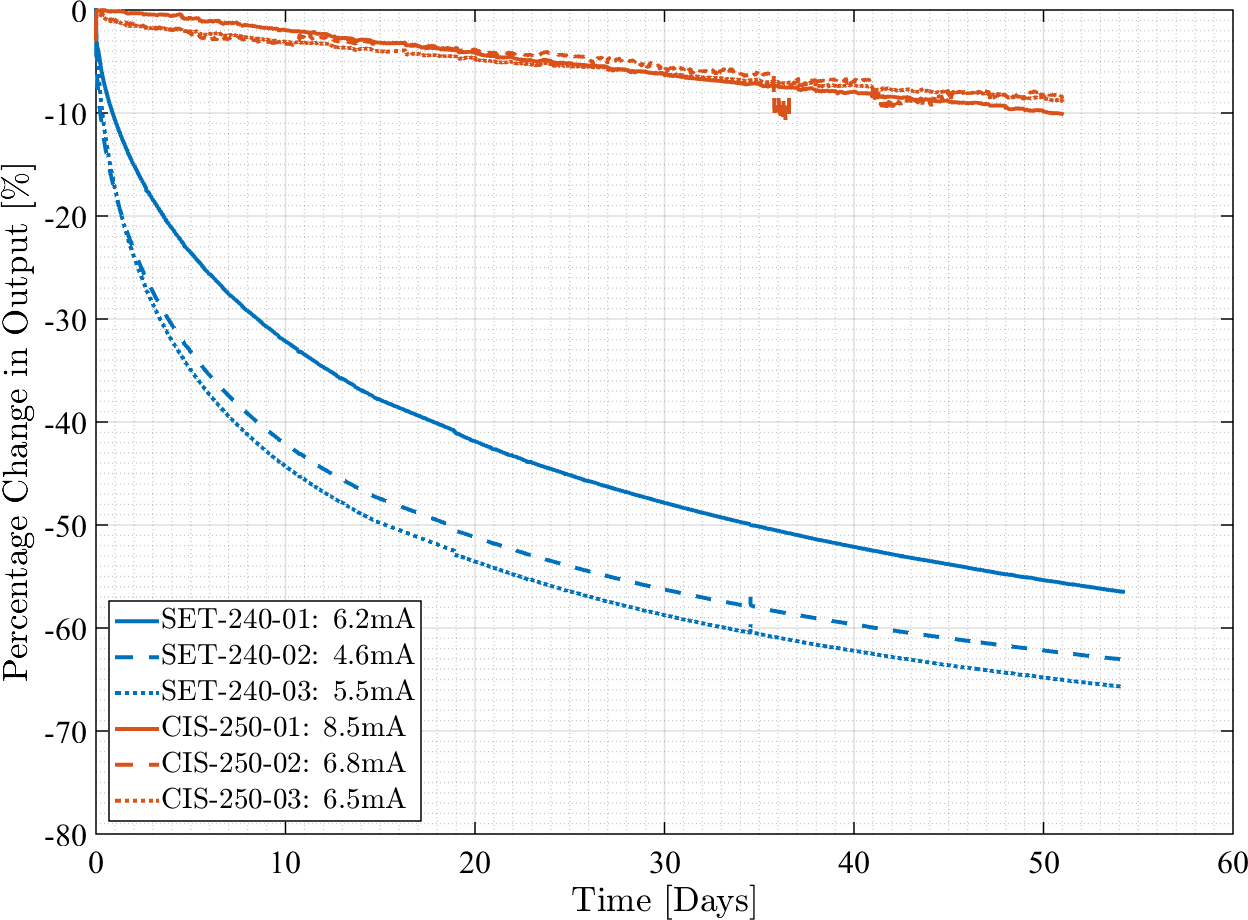}
\hfil
\includegraphics[width=0.32\textheight]{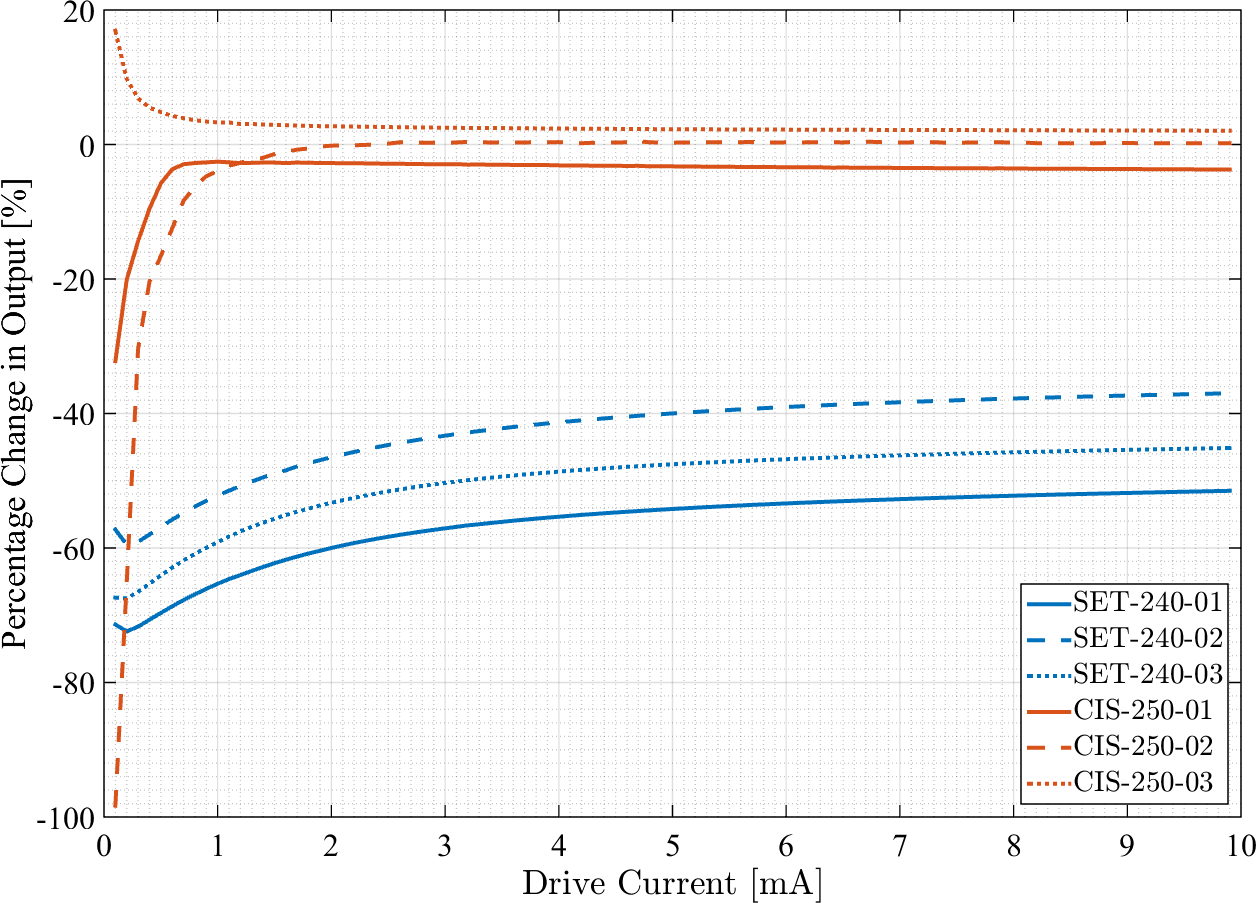}
\caption[Change in UV output.]{\label{fig:DCFDchangeUV} Left: The normalised UV output for each device measured in real-time over the duration of the test. While the data were recorded at $1\,\textup{Hz}$, here an average output was calculated from the 240 seconds of each cycle a device was on. Note these data also contain any degradation of the monitoring photodiodes. Right: While simultaneously measuring their UV output, an IV scan was performed on each device before and after the DC fast discharge lifetime test. The percentage differences in UV output between the two scans are shown.}
\end{figure}

As can be seen in Figure \ref{fig:DCFDchangeUV}, over the duration of the 9000 cycles the real-time monitoring suggested the UV output of the SET-240 devices fell by between 57\% and 66\% while the CIS-250 devices fell by 8-10\%. However, the monitoring data also contains any degradation of the measurement photodiodes, for example via solarisation of their windows. For this reason a carefully controlled set of reference measurements were carried out with a different photodiode, before and after the lifetime test. The percentage change in each device's UV output at different drive currents are also shown in Figure \ref{fig:DCFDchangeUV}.

Based on the reference measurements, the CIS-250 devices showed less than a $\pm3\,\%$ change in output at drive currents above $1\,\textup{mA}$. However, previous trial runs had shown the reproducibility of the measurement itself was at a similar level due to the devices needing to be remounted between measurements. Therefore the CIS-250 measurements above $1\,\textup{mA}$ are consistent with no change in output. This suggests most of the measured real-time degradation was actually in the photodiodes. The SET-240 devices produced between $37\,\%$ (highest settings) and $72\,\%$ (lowest settings) less UV light by the end of the test with a systematic offset between the three devices being proportional to the average electrical power, the higher the drive current amplitude the faster the degradation. The reference measurements for the SET-240s suggest that up to $20\,\%$ of the degradation observed in real-time actually came from the monitoring photodiodes.

A calibrated measurement of each device's UV output spectrum was also performed before and after the lifetime test. Within the uncertainty of the measurement, no device showed a change in peak position ($\pm0.1\,\textup{nm}$) or FWHM ($\pm0.2\,\textup{nm}$). Interestingly, the waveform reference measurement made at $100\,\textup{kHz}$, $10\%$ duty cycle with a $10\,\textup{mA}$ drive current amplitude did show a change for the SET-240 devices but not for the CIS-250s. The rise times (10\% to 90\% voltage amplitude) for all three fell by approximately 12\%, going from on average $192\pm3\,\textup{ns}$ before the lifetime test to $168\pm1\,\textup{ns}$ after.

A DC fast discharge would likely only be required for LISA every two or three weeks and at the UV output levels chosen for testing could be comfortably achieved within the tested 240 seconds. Even if one envisaged a daily DC fast discharge the tested 9000 cycles represents a significant over-test being equivalent to a mission lasting almost 25 years. As such both device types tested fulfil the light source requirements within a DC fast discharge scenario with considerable margin.

\subsection{Pulsed Fast Discharge}

This test was designed to simulate a pulsed fast discharge scenario for a mission lasting multiple years. Like the DC fast discharge, a pulsed fast discharge would be carried out in a few minutes with the device then turned off for an extended period before the next fast discharge was required. For our pulsed fast discharge test we removed these rest periods completely to reduce the overall testing time. The devices were operated at a variety of stressed ambient temperatures in order to examine how this affected the degradation rate.

\begin{figure}[h]
\centering
\includegraphics[width=0.32\textheight]{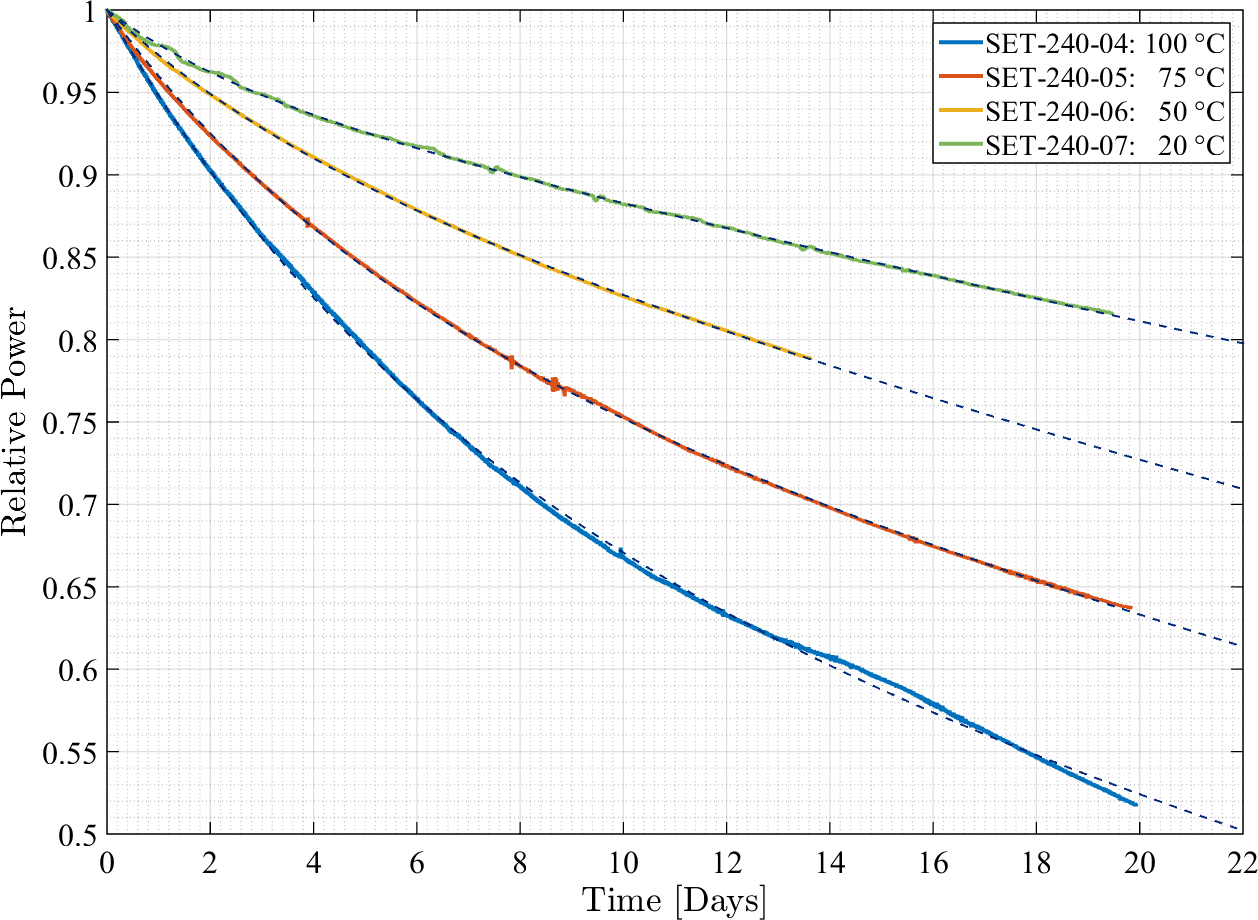}
\hfil
\includegraphics[width=0.32\textheight]{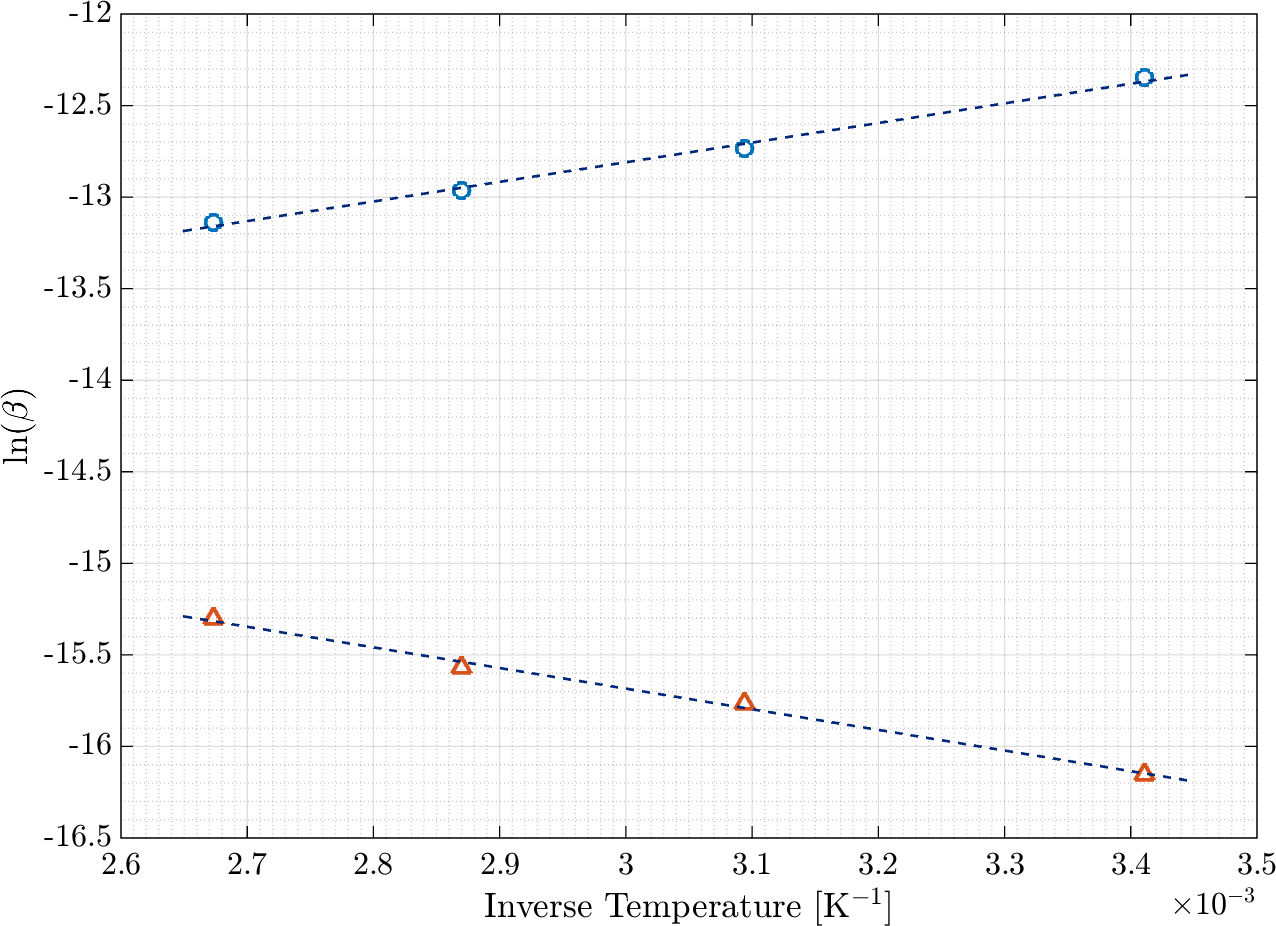}
\caption[Pulsed fast discharge lifetime test.]{\label{fig:Acc_LT} Left: The relative output of each device over the duration of the test with the fit to each degradation curve also shown. Note that each UV LED showed some initial device specific variation in output that clearly did not belong to a general trend. These fluctuations were at around the 5\% level and lasted less than a day for three of the devices but almost a week for device 06. In each case this initial section of data has been removed and the analysis carried out on the remaining data. Right: The decay parameters obtained from the fits plotted against the inverse of operational temperature. $\beta_{1}$ is denoted by the red triangles (temperature dependent) and $\beta_{2}$ by the blue circles (bias dependent).}
\end{figure}

Given the potential advantages of using a shorter wavelength light source these tests focused on the SET-240 UV LEDs. Three brand new devices were used (labelled 04, 05, 06), with each being fully characterised prior to the start of the test with our standard set of reference measurements, see Figure \ref{fig:DeviceChar}. Using the Arrhenius method \cite{Yamakoshi1981, Egawa1996, Gong2006}, the aim was to find a relationship between device degradation rate and operating temperature. To achieve this a single device at a time was mounted in a temperature controlled copper block within a vacuum chamber at a pressure of less than $10^{-5}\,\textup{mbar}$. With the device under test held at a stressed temperature ($100\,^{\circ}\textup{C}$, $75\,^{\circ}\textup{C}$ or $50\,^{\circ}\textup{C}$) it was pulse driven at $100\,\textup{kHz}$, $10\%$ duty cycle with a $10\,\textup{mA}$ drive current amplitude. In a separate mount opposite the UV LED, a large area photodiode (Hamamatsu S1337-1010BQ) was used to continuously measure the device's UV output as it degraded over 21 days. To extend the analysis we included the first 21 days worth of data from the pulsed continuous discharge measurement (described in the next section) that was made at $20\,^{\circ}\textup{C}$ with the SET-240 device labelled 07. Figure \ref{fig:Acc_LT} shows how each device's UV output declined over the course of the test. As expected, the rate of degradation increased with increasing temperature with a clear trend visible.

To analyse the data we took a similar approach to that described in \cite{Gong2006}. A non-linear reduced $\chi^{2}$ minimisation fit was performed with each decay curve with a double exponential of the form $P(t) = a_{1}e^{-\beta_{1} t} + a_{2}e^{-\beta_{2} t}$, where $P(t)$ is the relative UV output power with time, $a_{n}$ describes the initial amplitude of each term and $\beta_{n}$ are degradation rates. We set $a_{2}=1-a_{1}$ such that only three parameters were fit. We assign $\beta_{1}$ as a temperature-dependent parameter and $\beta_{2}$ as a bias-dependent parameter, although as the voltage required to drive the fixed $10\,\textup{mA}$ pulses is inversely proportional to temperature, $\beta_{2}$ also includes an element of temperature dependence. The two sets of degradation rate are shown in Figure \ref{fig:Acc_LT}.

For the degradation rate associated with temperature one can extract an activation energy for the degradation process by considering $\beta_{1} = a_{0}e^{\frac{-E_{\alpha}}{k_{B} T}}$, where $\beta_{1}$ is the degradation rate, $a_{0}$ is a constant, $E_{\alpha}$ is the activation energy, $k_{B}$ is the Boltzmann constant and $T$ is the temperature in Kelvin. An activation energy of $97\pm5\,\textup{meV}$ was obtained from the linear fit shown in Figure \ref{fig:Acc_LT}, in good agreement with literature values for similar devices that range from $100-270\,\textup{meV}$, \cite{Xi2005, Shatalov2006}.

The current-voltage scans that were part of the post-test reference measurements revealed a fall in each device's UV output at a DC drive current of $10\,\textup{mA}$ that were consistent with the degradation measured \textit{in situ} with the devices pulsed. The same measurements revealed no change in any of the device's spectral properties. However, as before the rise times of all four devices had fallen by a similar amount when driven at $100\,\textup{kHz}$, $10\%$ duty cycle with a $10\,\textup{mA}$ amplitude. Prior to the test the devices had rise times of $198\pm1\,\textup{ns}$, with all four falling to $171\pm2\,\textup{ns}$ by the end of the test. It is also noteworthy that while commissioning the test set up a run with a single CIS-250 device at $100\,^{\circ}\textup{C}$ was carried out. This test lasted four days but the device's output only fell by around 3\%. Combined with the DC fast discharge test results, this suggests they are extremely resilient.

At the drive settings used here ($100\,\textup{kHz}$, $10\%$ duty cycle with a $10\,\textup{mA}$ drive current amplitude) the average UV output power is approximately $12\,\mu\textup{W}$ at start of life. Considering conservative surface properties and fibre transmissions as before, it would be possible to discharge $10^{7}\,\textup{elementary charges}$ in a matter of minutes with such a pulsed fast discharge likely only being required every two or three weeks. Therefore the 21 days of continuous operation demonstrated here is equivalent to several tens of years of real-world use in a pulsed fast discharge mode of operation.

\subsection{Pulsed Continuous Discharge}

For our long-term pulsed lifetime study, a range of device drive settings, spanning those required for a fast pulsed discharge to those necessary for a continuous discharge scenario were examined. Based on the results of our pulsed fast discharge test, drive settings of $100\,\textup{kHz}$, $10\%$ duty cycle with a $10\,\textup{mA}$ amplitude were chosen to provide a comparison with the elevated temperature tests. However, continuous discharging would require photo-currents around a factor $10^{3}$ smaller than these settings would produce. There are several possibilities for achieving this with pros and cons in each case. Perhaps the most obvious would be to reduce the drive current amplitude to reduce the number of photons in each pulse. However, previous tests have shown that at drive currents less than $10\,\textup{mA}$ rise times compatible with $\sim1\,\mu\textup{s}$ pulses may be difficult to achieve, \cite{Hollington2015}. Shifting the phase of the UV pulse with respect to the $100\,\textup{kHz}$ bias within the sensor until the direction of the instantaneous fields produces the desired photo-current direction is also an option. While having some ability to shift the phase would be highly desirable the exact relationship between phase and net photo-current would be complicated.

The concept settled on to reduce the UV power output was to use $10\,\textup{mA}$ amplitude pulses lasting $1\,\mu\textup{s}$, but vary the repetition rate. This could be achieved by counting the number of $100\,\textup{kHz}$ cycles and  only generating a $1\,\mu\textup{s}$ pulse on the $n$th cycle. This method offers a high dynamic range and may be a relatively simple yet elegant way of driving devices in a flight like system.

For the three devices under test $n=1,\,33\,\textup{and}\,1000$ were used, which are referred to as High, Medium and Low output levels respectively. The High level is equivalent to the pulsed fast discharge, $100\,\textup{kHz}$, $10\%$ duty cycle with a $10\,\textup{mA}$ amplitude. For SET-240-07, this gave an initial average UV output power of $\sim12\,\mu\textup{W}$. The Medium level is equivalent to $3\,\textup{kHz}$, $0.3\%$ duty cycle with a $10\,\textup{mA}$ amplitude. For SET-240-08, this gave an initial average UV output power of $\sim420\,\textup{nW}$ or $\approx31.6$ times less UV power than the previous setting. The Low level is equivalent to $100\,\textup{Hz}$, $0.01\%$ duty cycle with a $10\,\textup{mA}$ amplitude, giving an initial UV output power for SET-240-09 of $\sim13\,\textup{nW}$ or $\approx31.6$ times less UV power than the previous setting. So simply counting $100\,\textup{kHz}$ cycles gives a dynamic range of $\approx1000$, with average UV output powers at the High level appropriate for pulsed fast discharging while at the Low level those suitable for continuous discharging.

\begin{figure}[h]
\centering
\includegraphics[width=0.32\textheight]{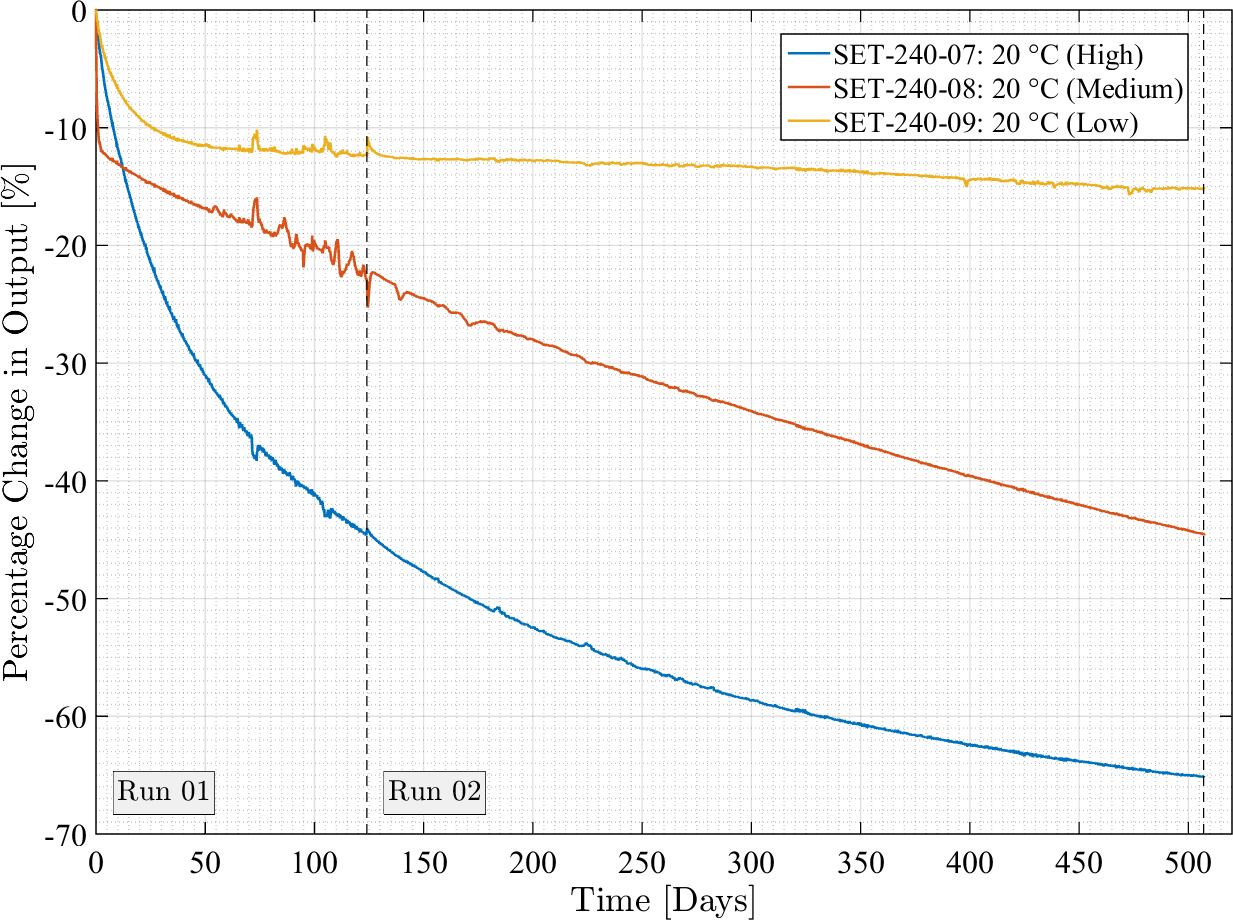}
\hfil
\includegraphics[width=0.32\textheight]{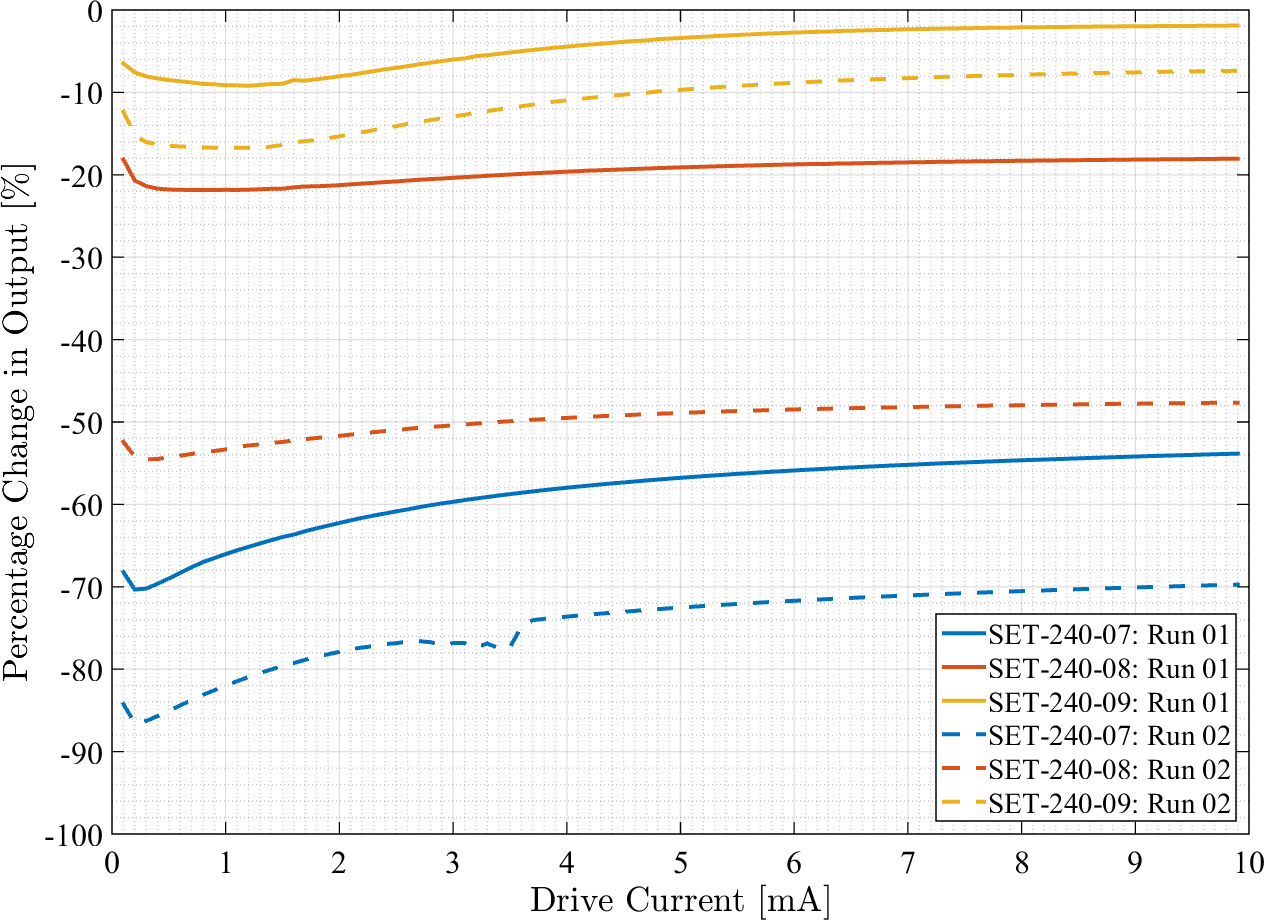}
\caption[Change in UV output during long-term lifetime test.]{\label{fig:PulsedLongTermUV} Left: The percentage change in each device's output, as measured by the real-time monitoring. Right: Based on the dedicated reference measurements, one can see an independent measurement of the degradation of each device at the end of Run 01 and again at the end of Run 02.}
\end{figure}

The equipment used for this lifetime test was the same as that used for the DC fast discharge, with the devices held at $20\,^{\circ}\textup{C}$ and each device's average output readout by a separate photodiode at $10^{-1}\,\textup{Hz}$. The test ran for a total of 507 days which was split into two runs in order to allow a full set of reference measurements of be made after 124 days of operation. The results are shown in Figure \ref{fig:PulsedLongTermUV}.

Unsurprisingly, the real-time monitoring showed the overall level of degradation increased with increasing repetition rate, falling $66\%$, $45\%$ and $15\%$ compared to their starting values respectively. However, a closer look at each degradation curve reveals some significant differences between each device. While SET-240-07 showed a fairly smooth decay, the behaviour of the other two devices appears to have varied as the test progressed. This is particularly clear for SET-240-08, whose output rapidly declined by around $10\%$ over the first few days before the degradation rate abruptly changes, becoming almost constant for the remainder of the test. Another feature of the data is the apparent erratic behaviour during the second half of Run 01, which was seen in all devices but particularly clearly for SET-240-08. While the copper block the devices were mounted in was independently held at $20\,^{\circ}\textup{C}$, a temporary problem with the laboratory temperature control affected the drive and readout electronics leading to these transient anomalies with the issue resolved for Run02.

Figure \ref{fig:PulsedLongTermUV} also shows the change in each device's UV output obtained from our standard reference measurements made at the end of each run, made with a dedicated photodiode which otherwise remained unused in safe storage. These DC measurements should be superior to those carried out in real-time which are sensitive to any photodiode degradation as well as any change in device rise time. Reassuringly, these measurements show a similar level of degradation at $10\,\textup{mA}$ for each device compared to the real-time measurements, equivalent to falls of $70\%$, $48\%$ and $8\%$ of their initial levels by the end of Run 02.

Similar to the other lifetime tests, the device reference measurements showed no change in spectral properties but a fall in rise times. Prior to the test the three devices all had 10\% to 90\% rise times of $198\pm1\,\textup{ns}$. Going from High to Low, after Run 01 they had fallen by 7.5\%, 3.4\% and 0.9\% respectively and by the end of Run 02 had fallen a total of 8.7\%, 6.3\% and 2.1\%. Although the overall changes are quite small the UV pulses are nominally $1\,\mu\textup{s}$ in duration so a rise time of $\approx0.2\,\mu\textup{s}$ is significant and produces a measurable increase in the average UV output when the rise time falls. In the real-time monitoring where the devices were pulsed, this fall in rise time increases the average UV output acting against the underlying decay. This goes some way to explaining the slight discrepancy between the fall in output measured in real-time compared to those based on the reference measurements.

This long-term study also shows that the double exponential model derived with the pulsed fast discharge test underestimates the long-term lifetime. After 507 days of operation at $20\,^{\circ}\textup{C}$ while driven at $100\,\textup{kHz}$, $10\%$ duty cycle with a $10\,\textup{mA}$ amplitude, the Arrhenius analysis predicted that the UV output of a SET-240 device would fall by 98\%. However, the single device that was operated at these settings during the long-term lifetime test only fell by $\sim65\%$. At these drive settings the long-term study goes far beyond the lifetimes required for pulsed fast discharging but nonetheless it demonstrates the need for care when extrapolating long-term behaviour.

Over the course of the 507-day test, the device driven at the low level suitable for continuous discharging experienced a fall in output of around 10\%, the majority of which occurred in the first 50 days. Barring a sudden complete failure of a device, the degradation rate seen at these settings would be compatible with several years of continuous discharging. The likely inclusion of redundant devices would obviously extend the lifetime of such a system still further.


\section{Conclusions}

We performed a series of lifetime tests at output levels suitable for both fast and continuous discharging in both DC and pulsed modes of operation. We started with a DC fast discharge test. Spanning 50 days it represented over 9000 fast discharge cycles. After accounting for any photodiode degradation, the three CIS-250 devices tested showed less than a $4\,\%$ change in UV output at DC drive currents above $1\,\textup{mA}$. Meanwhile the SET-240 devices showed a significant fall in UV output ranging from a $-40\,\%$ to $-70\,\%$. However, none failed and this represented a significant over-test in terms of both the UV output level chosen and the number of fast discharge cycles, which were equivalent to a mission lasting almost 25 years.

Focusing on the SET-240 devices which provide shorter wavelength UV light, we performed further lifetime tests. The pulsed fast discharge test lasted 21 days and demonstrated the performance of the SET-240 at stressed temperatures of up to $100\,^{\circ}\textup{C}$. While the UV output of the devices fell significantly, at the drive settings chosen the usage represented several tens of years worth of fast discharges at extreme temperatures.

Finally, the long-term pulsed lifetime test represented output levels for the SET-240 suitable for both fast and continuous pulsed discharge. Overall, the results showed that the devices can be operated continually in a pulsed mode on time-scales comparable to a multi-year LISA mission. In particular, at the low repetition rate suitable for continuous discharging, the average UV output of the device fell by only $\sim10\%$ over 507 days.

Some general observations about the ageing behaviour of the SET-240s can also be made. For example, though some significant falls in UV output were observed, out of 9 devices, none completely failed during any of the tests. They were also all found to be very spectrally stable with age but were consistently found to show a decrease in rise times. Such behaviour may need to be considered in a full system as it determines the average UV output during a pulse. 

We also found a certain amount of variation between devices with respect to their ageing behaviour. As one may have expected, this seemed clearest early in a device's life before eventually setting down. For use in space it would likely be wise to take a larger group of devices than required and put them through a controlled burn-in period prior to selecting the final ones for use. This would also flag early any defective devices and offer more predictable behaviour for those chosen. In addition to a device's output being strongly temperature dependent, we have also seen how important thermal management is to maximising a device's lifetime. While active temperature control is likely an unnecessary complication in a future flight system, ensuring suitable heat sinking will be vital.

The LISA charge management system will need to be capable of performing fast discharges every two or three weeks and ideally also be capable of discharging continuously without introducing excess noise to the primary measurement. Whether these aims are best achieved with a DC or pulsed mode of operation is still open for debate as the performance of the CMS is tightly entwined with the photoelectric properties of the sensor surfaces. Our understanding of these will be the strongest driver in the design of the final flight system. Ultimately there is a trade-off between the simplicity of a DC system which relies on the sensor having predicable photoelectric properties, or a more complex pulsed system which has superior robustness for a sensor with unknown photoelectric properties.

Whichever route is taken, all tests have shown that the SET-240 devices would make an excellent UV light source for a LISA mission lasting five or more years, with the CIS-250 devices providing a reliable backup, though with the caveat of them having a longer peak wavelength.


\section*{Acknowledgements}

The authors would like to thank the whole LTPDA software development team \cite{LTPDA2015} as well as Shahid Hanif at Imperial College for designing and building the pulsed drive electronics. We also acknowledge the financial support of the European Space Agency under contract C4000103768 for this work.


\section*{References}

\bibliographystyle{unsrt}
\bibliography{bibliography}

\end{document}